\documentclass[preprint]{aastex}

\RequirePackage{xspace}

\def\ifundefined#1{\expandafter\ifx\csname#1\endcsname\relax}

\ifundefined{ensuremath}\def\ensuremath#1{\relax\ifmmode{#1}}
\else${#1}$\fi\else\relax\fi
\ifundefined{nuc}\def\nuc#1#2{\relax\ifmmode{}^{#1}{\protect\mathrm{#2}}
\else${}^{#1}$#2\fi}\else\relax\fi
\newcommand{\etal}{et al.\xspace}
\newcommand{\gcm}{g~cm$^{-3}$\xspace}
\newcommand{\kmps}{\ensuremath{\mathrm{km}~\mathrm{s}^{-1}}\xspace}

\newcommand{\msol}{\ensuremath{{\mathrm{M}_\odot}}\xspace}

\newcommand{\nni}{\ensuremath{\nuc{56}{Ni}}\xspace}

\newcommand{\phx}{\texttt{PHOENIX}\xspace}
\newcommand\phxO{\texttt{PHOENIX/1D}\xspace}

\newcommand{\gamray}{$\gamma$-ray\xspace}

\newcommand{\mch}{M\ensuremath{_\mathrm{Ch}}}
\newcommand{\mni}{M\ensuremath{_\mathrm{Ni}}}

\newcommand{\snooneay}{SN~2001ay\xspace}
\newcommand{\dmphil}{\ensuremath{\Delta m_{15}}\xspace}
\newcommand{\bmv}{\ensuremath{B\,-\,V}\xspace}

\graphicspath{{./}{./figures/}{../talk/trieste11/figures/}}

\begin{document}

\shorttitle{A Physical Model for \snooneay}
\shortauthors{Baron, H{\"o}flich, \etal}

\title{A Physical Model for \snooneay, a normal, bright, extremely slow
  declining Type Ia supernova}

\author{E. Baron\altaffilmark{1,2,3,4}, 
  P. H\"oflich\altaffilmark{5}, 
  K. Krisciunas\altaffilmark{6}, 
I. Dominguez\altaffilmark{7},    A. M. Khokhlov\altaffilmark{8},
M. M. Phillips\altaffilmark{9},  N. Suntzeff\altaffilmark{6}, 
  L. Wang\altaffilmark{6} 
  }

\altaffiltext{1}{Homer L.~Dodge Department of Physics and Astronomy,
  University of Oklahoma, 440 W.~Brooks, Rm 100, Norman, OK,
  73019-2061 USA; baron@ou.edu}
\altaffiltext{2}{Hamburger Sternwarte, Gojenbergsweg 112, 21029
  Hamburg, Germany}
\altaffiltext{3}{Computational Research Division, Lawrence Berkeley
        National Laboratory, MS 50B-4206, 1 Cyclotron Rd, Berkeley, CA
        94720 USA}
\altaffiltext{4}{Physics Department, University of California,
        Berkeley, CA        94720 USA}

\altaffiltext{5}{Department of
  Physics, Florida State University, Tallahassee, FL 32306, USA;
  pah@astro.physics.fsu.edu} 

\altaffiltext{6}{George P. and Cynthia
Woods Mitchell Institute for Fundamental Physics \& Astronomy,
Texas A \& M University, Department of Physics \& Astronomy,
4242 TAMU, College Station, TX 77843, USA;
  krisciunas@physics.tamu.edu, suntzeff@physics.tamu.edu,
  lwang@physics.tamu.edu} 

\altaffiltext{7}{Universidad de Granada
C/ Bajo de Huetor 24 Aptdo 3004, ES 18071, Granada, Spain; inma@ugr.es}

\altaffiltext{8}{Department of Astronomy and
  Astrophysics, University of Chicago, Chicago, IL, USA;
  ajk@oddjob.uchicago.edu}

\altaffiltext{9}{Las Campanas Observatory, Casilla 601, La Serena,
  Chile; mmp@lcoeps1.lco.cl}

\begin{abstract}
  We present a study of the peculiar Type Ia supernova 2001ay
  (\snooneay). The defining features of its peculiarity are: high
  velocity, broad lines, and a fast rising light curve, combined with
  the slowest known rate of decline. It is one magnitude dimmer than
  would be predicted from its observed \dmphil, and shows broad
  spectral features.  We base our analysis on detailed calculations
  for the explosion, light curves, and spectra. We demonstrate that
  consistency is key for both validating the models and probing the
  underlying physics.  We show that this SN can be understood within
  the physics underlying the \dmphil relation, and in the framework of
  pulsating delayed detonation models originating from a Chandrasekhar
  mass, $\mch$, white dwarf, but with a progenitor core
  composed of 80\% carbon.  We suggest a possible scenario for stellar
  evolution which leads to such a progenitor.  We show that the
  unusual light curve decline can be understood with the same physics
  as has been used to understand the \dmphil--relation for normal
  SNe~Ia. The decline relation can be explained by a combination of
  the temperature dependence of the opacity and excess or deficit of
  the peak luminosity, $\alpha$, measured relative to the instantaneous
  rate of radiative decay energy generation.  What differentiates
  \snooneay from normal SNe~Ia is a higher explosion energy which
  leads to a shift of the \nni distribution towards higher velocity and
  $\alpha < 1$. This result is responsible for the fast
  rise and slow decline.  We define a class of SN~2001ay-like SNe~Ia,
  which will show an anti-Phillips relation.
\end{abstract}


\section{Introduction}\label{Introduction}

While some progress has been made in the understanding of the Type Ia
supernova (SN~Ia) phenomenon in recent years, there does not yet exist
an agreed upon standard model of the supernova explosion that can
explain normal SNe~Ia \citep{branchcomp105,branchcomp509}.  For
cosmology, the brightness decline relation plays a key role
\citep{philm15,philetal99,goldhetal01}.  From theory, $\dmphil$ is
well understood: the light curves (LCs) are powered by radioactive decay of \nni
\citep{ColgateMckee69}.  More \nni increases the luminosity and causes
the envelopes to be hotter. Higher temperature means higher opacity
and, thus, longer diffusion time scales and slower decline rates after
maximum light \citep{HKWPSH96,nughydro97,umeda99,kasen09}. The
existence of a $\dmphil$--relation holds for virtually all scenarios
as long as there is an excess amount of stored energy to be released
\citep{HKWPSH96}. Although, the tightness of the relation can be
understood within the framework of the single degenerate scenario and
spherical models \citep{HKWPSH96,HGFS99by02,hoeflich10}, it falls
apart when taking into account burning instabilities during the
deflagration phase \citep{kasen09}.  This difficulty and the
observation of a set of extremely bright SNe Ia may lend support for
double degenerate scenarios with progenitors well above the
Chandrasekhar mass
\citep{howell03fg06,scalzo07if10,tauben09dc11,howell11_rev}. We note,
however, that the inferred brightness depends on a unique relation
between the \nni mass $\mni$, and the intrinsic color: \bmv at maximum
light. At least in a few cases, the apparent brightness can be
understood within the framework of $\mch$ mass WDs with intrinsically
red color \citep{quimby05cg06}.

Additionally, some progress has been made in understanding variations
among the groups, with suggestions that some of the spectral diversity
is due to metallicity, central density, and asymmetries
\citep{hwt98,lentzmet00,hoeflich10,maedanature10,maund10a}.  The
nature of the progenitor system is also controversial, with recent
work on rates and the delay time distribution favoring the double
degenerate scenario,
\citep{yl98,YL00,maoz08,ruiter09,maoz10a,mennekens10,ruiter11,maoz11a}. Nevertheless, 
theoretical work continues to favor the single degenerate scenario,
with some contribution of double degenerate scenario
\citep{hk96,sainom98,ww86_mergers,moch_livio90,sainom85,shen_mergers11}.

Despite the uncertainties in theory, most of the known SNe~Ia obey the
brightness decline relation. The light curves are self-similar within
$\pm 0.3$~mag \citep{goldh98,riessetalIaev99}.  Deviations of this
order can be expected from variations of the progenitor
\citep{hwt98,brach00,Thiel03,sitenzahl11}. 
The stretching method works for both the local and
high-z samples \citep{perletal99,riess_scoop98}. The self-similarity
holds even for supernovae for which super-Chandrasekhar mass
progenitors have been suggested.

The subject of this paper, \snooneay, shows that nature is even more
diverse \citep{krisc01ay11}.  From the redshift of the
host galaxy IC4423  \citep{WFfinal01},
the distance modulus $\mu = 35.55 \pm 0.1$ is well known. The
reddening in our galaxy is found to be $\mathrm{E(B-V)} = 0.026 \pm
0.006$~mag, and interstellar absorption in Na~I suggests low reddening
in the host galaxy of about $\mathrm{E(B-V)} = 0.072 \pm 0.008$~mag
\citep{krisc01ay11}.  Together with the observed magnitude, the
intrinsic brightness at maximum light is $M_V = -19.17$~mag with a \bmv color
of $-0.02$~mag, which is comparable with typical core-normal
supernovae.

However, the light curve shape is unlike other SNe~Ia: Its measured
decline rate of $\Delta m_B= 0.68$, is slower than any known SNe~Ia,
combined with a fast rise of some 16 days \citep{krisc01ay11}. Based
on its \dmphil, \snooneay should be brighter than observed by about $1$~mag. Moreover,
the linewidths near maximum light put it solidly in the broad-line
class of SNe~Ia.

We show that \snooneay can be understood with the same physics
underlying the \dmphil--relation, and in the framework of
parametrized pulsating delayed detonation models similar to SN1990N
\citep{kmh92}, but with an unusual progenitor star.  We treat
consistently the explosion, light curves, and spectra
\citep{h95}. We show that consistent models reproduce the observed
light curves and spectra reasonably well. Furthermore, we show that inconsistent
calculations lead to spectral features which would lead to rejection
of the explosion model.  Finally, we summarize
our findings and discuss possible implications for the understanding
of SNe~Ia and cosmology.

\section{Motivation for our Model for \snooneay}
\label{lc_dis}

For our models, an understanding of the brightness decline relation is
important. The Phillips relation \citep{philm15,philetal99}
provides an empirical link between peak brightness and
decline of the LC after maximum. Most SNe~Ia fall within 0.2~mag of
this relation. 

As discussed in the introduction, the $\dmphil$--relation can be
understood as an opacity effect if energy in stored energy is
available in excess of the instantaneous radioactive decay
\citep{HKWPSH96,nughydro97,umeda99,kasen09}.  The latter is a key to
understand \snooneay.  We will discuss that while the second condition
is valid in most scenarios for SNe~Ia, it is not guaranteed.

For our discussion a useful quantity is the relation between instantaneous
energy deposition by radioactive decay and the brightness at maximum:
\[ L_{\mathrm{bol}}(t_{\mathrm{max}}) = \alpha \dot
S(t_{\mathrm{max}}) \] where $\alpha$ accounts for the fact that some
of the \gamray energy deposited prior to maximum light can be stored
in the thermal energy and trapped radiation energy which cannot escape
faster than a diffusion time. In a typical delayed detonation model
the value of $\alpha$ is about $1.2$ \citep{kmh93,hk96}.

The role of the opacity condition can be seen from Arnett's analytic
solution \citep{arnett80,arnett82}, which also provides further insights.
Constant opacity in a polytropic model with homogeneous energy input
does not produce a brightness decline relation. In this model, energy
balance between radioactive decay and adiabatic expansion and cooling
cancel exactly in a radiation dominated regime with pure geometrical
dilution. At maximum, the ratio between maximum brightness and instantaneous
energy input is $\alpha = 1$.  \citet{PE00} reconsidered Arnett's one-zone
models and confirmed previous explanations.  Declining opacity provide
extra energy at maximum light because it accelerates the recession of
the ``photosphere'' \citep{hmk93b} and results in a positive brightness
decline relation in this case.

However, even an effective opacity declining with temperature may
result in an anti-$\dmphil$--relation, because $\dmphil$ depends both
on the opacity effect and the excess energy available.
For example, the pure detonation model DET1 has an $\alpha \approx
0.73$ and \dmphil of $1.37$~mag.  In contrast, another pure detonation
model DET2 has an $\alpha \approx 1.2$ and $\dmphil \approx 1.7$
\citep{kmh93,hk96}. In DET1, the high central density leads to
significant electron capture in the central region producing a \nni
free core. In contrast, lower central densities in DET2 produce \nni
in the center, and less \nni in the outer region. In both models, the
density structure is close to similar polytropes but the \nni
distribution is shifted outwards and inwards, respectively.  The
differences in $\alpha $ can be understood in terms of the deviation
from Arnett's one-zone model with a flat \nni distribution in mass. In
DET1, this distribution is shifted outwards leading to an increased
escape probability of \gamray{s} and greater expansion work which
leads to  reduced
$\alpha$. In DET2, the shift of the \nni distribution inwards leads to
a larger value of $\alpha$.

From the above discussion, it is obvious that a
narrow \dmphil-relation requires similar abundance patterns and \nni
distributions for a given $\mni$. The theoretical relation $\dmphil
(\mni)$ depends on the explosion scenario.  Self-similarity in the
light curves requires a self-similar transformation between models of
differing brightness.  Within the delayed detonation scenario, $\mni$
depends mostly on the pre-expansion during the deflagration phase,
since if the density is too high neutron-rich iron-group elements will
be produced. However, as the nickel mass drops out of the standard
0.5--0.6~\msol range, 
the \nni distribution shifts inwards with decreasing $\mni$.  This
produces a $\dmphil (\mni)$-relation which is in agreement with observations
\citep{HGFS99by02}.  For core-normal SNe~Ia, the temperature and, with
it, the opacity remains high well after maximum light, leading to a
``shallow'' $\dmphil (\mni)$. Below a certain $\mni$, the nickel is
very centrally condensed (there is no nickel hole) and the opacity
drops rapidly soon after maximum leading to a steep $\dmphil (\mni)$-relation. 
In the case of fast decliners \nni is only produced during the deflagration
phase.

\section{Scenarios for \snooneay}

\subsection{Models with an increased mass}

\snooneay was suggested to fall within the class of ``Super-Chandra''
mass SNe~Ia.  Obviously, \snooneay\ does not obey the standard
\dmphil--relation. Within the framework of the empirical
\dmphil--relation, \snooneay\ should be brighter than observed by
roughly 1.0 magnitude which can be ruled out \citep[see Fig. 10
of][]{krisc01ay11}.  In fact, compared to the putative
super-Chandrasekhar SNe~Ia, 2007if \citep{scalzo07if10,yuan07if10}
and SN 2009dc
\citep{yamanaka09dc09,tanaka09dc10,silverman09dc11,tauben09dc11}
estimates of  $M_{\mathrm{bol}}$ at maximum light differ by about 1
magnitude. Estimates for the brightness of \snooneay imply 0.5 $\mni$
\citep{krisc01ay11}.  We need an increase of the diffusion time
scales by a factor of 2 from \dmphil\ of $1.25$ for typical SNe~Ia
\citep{philetal99} to $\dmphil = 0.68$.  The diffusion times scale as,
$t_\mathrm{diff} \propto \tau^2 \propto M^2 $ and, for homologous
expansion, $\tau \propto t^{-2}$.  If we assume $\mch$ for normal
SNe~Ia, this would imply a progenitor mass of $ 2 \msol$ which is
super-Chandra and in fact typical of the mass suggested for mergers,
but then one must explain why some super-Chandras are extra bright and
\snooneay is not.

A stronger argument against a high mass progenitor comes from the rise
time to decline ratio. \snooneay 
rises quickly to maximum, whereas an increased time scale would
unavoidably produce a slow rise to maximum light.

\subsection{The case for an $\mch$ mass WD with a faster expansion rate}

Here, we suggest another scenario outside the ``classical''
regime for SNe~Ia, but within the delayed-detonation scenario of
Chandrasekhar mass white dwarf progenitors. To reproduce \snooneay,
we require a fast rising light curve followed by a decline slower than
the slope of the instantaneous radioactive decay. We can produce
models with $\alpha < 1 $. This requires a model with \nni shifted to high 
velocities.  Since eventually, the luminosity is given by the rate of instantaneous energy
deposition, such a model will show a slow decline by construction.

\section{Results}

\subsection{Explosion Models and Light Curves}

We describe how a change of parameters can transform the results
from that given by normal LCs to that of \snooneay, within the same
physical picture.

Even in the absence of a well agreed upon model for SNe~Ia most of the
basic observational properties, that is, light curves and spectra, of core normal
SNe~Ia can be understood within the framework of thermonuclear
explosions of Chandrasekhar mass white dwarfs (WD) and, in particular, the
delayed-detonation scenario \citep{khok89} which provides a
natural explanation for the wide variety and range of the \nni\ 
production. To first order, the amount of \nni\  produced in the explosion depends on the
pre-expansion of the WD during the deflagration phase, which in
spherical models, can be parametrized conveniently by the density at
which the transition between deflagration to detonation occurs.  For a
recent review of the relation between observational properties and the
underlying physical model, see, for example, \citet{hoef_review07}.

We base our analysis on a spherical explosion model within the
framework of the pulsating delayed detonation scenario.  Our goal is
to construct a model with reduced $\alpha$ and short diffusion time
scales.  We seek to obtain our goals by increasing the expansion ratio
and shifting of the \nni distribution to higher velocity.

\subsubsection {Explosion Models}
\label{explosion}
 
The spherical explosion model has been constructed to allow fits of
optical light curves and spectra of \snooneay.
 
Within the DD scenario, the free model parameters are: 1) The chemical
structure of the exploding WD, 2) Its central density, $\rho_c$ at the
time of the explosion, 3) The description of the deflagration front,
and 4) the layer at which the transition from deflagration to
detonation occurs.

As reference, we started from the delayed detonation model
5p0y25z22.25, which has been found to be a good starting point for
core-normal SNe~Ia,  with respect to both spectra and light curves
\citep{DHS01a,h02,quimby05cg06,hoeflich10}. This base model originates
from a star with a main sequence mass of 5 \msol and solar
metallicity.  Through accretion, the C+O core of the star has grown close to the
Chandrasekhar limit.  At the time of the explosion of the WD, its
central density is 2.0$\times 10^9$ g~cm$^{-3}$ and its mass is close
to 1.37$M_\odot$. The transition density $\rho_{tr}$ has been
identified as the main factor which determines the \nni production
and, thus, the brightness of SNe~Ia
\citep{h95,hkw95,HGFS99by02,iwamoto99}.  The transition density
$\rho_{tr}$ from deflagration to detonation is $25 \times 10^6$
g~cm$^{-3}$.

 For \snooneay, we tuned the parameters.  The reference model has been
 modified as follows: We reduced the central density, $\rho_c$ to $1
 \times 10^9$~\gcm, which  decreased the potential energy, as always, and  thus,
  increased the radius of 
 the WD by $\approx 30$\% to $2320$~km, and we increased the C/O ratio
 in the former He-burning core ($M < 0.56$~\msol) to 4.  Both modifications increase
 the explosion energy to 1.69 $\times$ 10$^{51}$ ergs (1.69 foe), 
leading to a more rapid expanding
 envelope, increasing the rate of geometrical dilution, and shifting
 the 
 \nni to higher velocity.  These effects are responsible for both the fast rise
 and slower decline of \snooneay.

In classical DD models, the deflagration to detonation transition,
DDT, occurs in a WD already unbound during the deflagration phase
\citep{khok89,yamaoka92}.  In contrast, in pulsating delayed models,
PDD, the WD remains bound after the deflagration phase, and the DDT
occurs during pulsation \citep{khok93,hkw95}.
 Other authors have suggested variations on the PDD
 \citep{ivanova74,bravo_prd06,bprd07,bravo_prd09a,bravo_prd09b}, but
 here when we discuss the PDD model, we specifically refer to the model
 of \citet{khok93}.
Reducing the 
burning rate during the deflagration phase moves the model from
the DD to the PDD regime in which the WD is bound at the end of the
deflagration phase. We artificially reduced the Atwood number, which we approximate as a
constant independent of density and composition, (see Appendix) from
0.2 to (0.14, 0.12, 0.10) for the series PDD\_11a--c which all undergo
weak pulsation --- producing LCs with rise times
between 14 and 16 days with a slow decline from maximum.  The best fit to
the observations is given by 
PDD\_11b.  Its details are described below.  While some of the
observed trends that we find with the PDD model may be obtained in a
3-D DD model including the effects of rotation, exploring such models
is beyond the scope of this work. We mention results from
PDD\_11a/c below when needed.

\begin{figure}
\includegraphics[scale=0.65,angle=0]{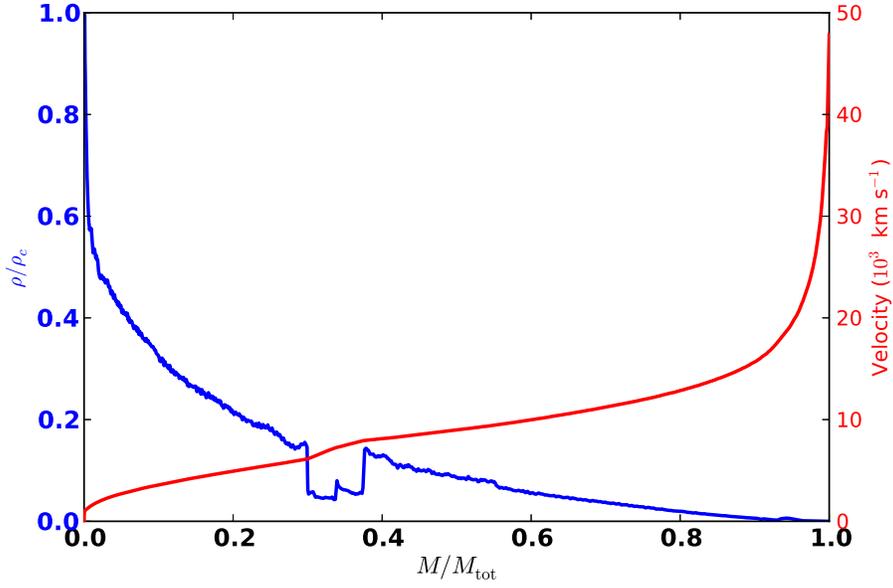}
\caption{Density and velocity as function of mass
for model PDD\_11b.} 
\label{fig:density}
\end{figure}

\begin{figure}
\includegraphics[scale=0.65,angle=0]{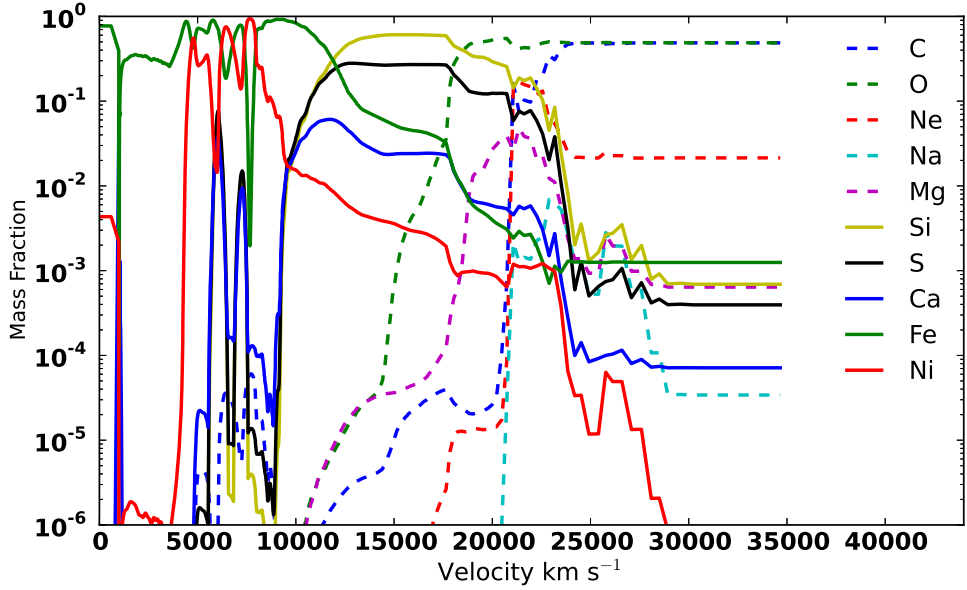}
\caption{Abundances of stable isotopes as a function of the expansion velocity
for model PDD\_11b.} 
\label{fig:abundance}
\end{figure}

 The density and abundance structure of PDD\_11b is given in
 Figs.~\ref{fig:density}--\ref{fig:abundance}.  The pulsation leaves a shell of $\approx 0.06
 \msol $ of unburned C/O (with C/O $\sim 1$, only the region of
 central helium burning has enhanced C/0), and $0.51 \msol$ of \nni.  For pulsating
 delayed detonation models, mixing during the pulsation may occur.  In
 contrast to previous pulsating models \citep{hkw95,hk96}, we didn't
 assume full mixing during the pulsation, but instead mixing was limited to one
 scale height from the position of the burning front during the
 pulsation. The mixed region between burned matter forms the layer
 where the detonation is born.  As a result, the density and chemical
 structure are similar to spherical DD models and consistent with late
 time line profiles and SN-remnants. We abandoned full mixing in order
 to preserve the inner, unmixed region as required by late time line
 profiles and the imaging of the supernova remnant S Andromedae
 \citep{hof03du04,moto06,fesen07,gerardy07,maedanature10}.  Our models
 leave some unburned C/O. All modifications combined result in broader
 lines. We note that our model is highly parametrized:
 Firstly, the deflagration phase depends critically on the initial
 condition and as discussed above, the problem of how to suppress strong
 mixing is still unsolved, though it may be related to high magnetic
 fields \citep{Penney11}. Second, the amount of mixing will depend 
 on Rayleigh-Taylor (RT) instabilities during the pulsation and possibly, rotation induced sheer
 instabilities.

\subsubsection{Light Curves}

\begin{figure}[t]
\begin{center}
\includegraphics[width=\textwidth]{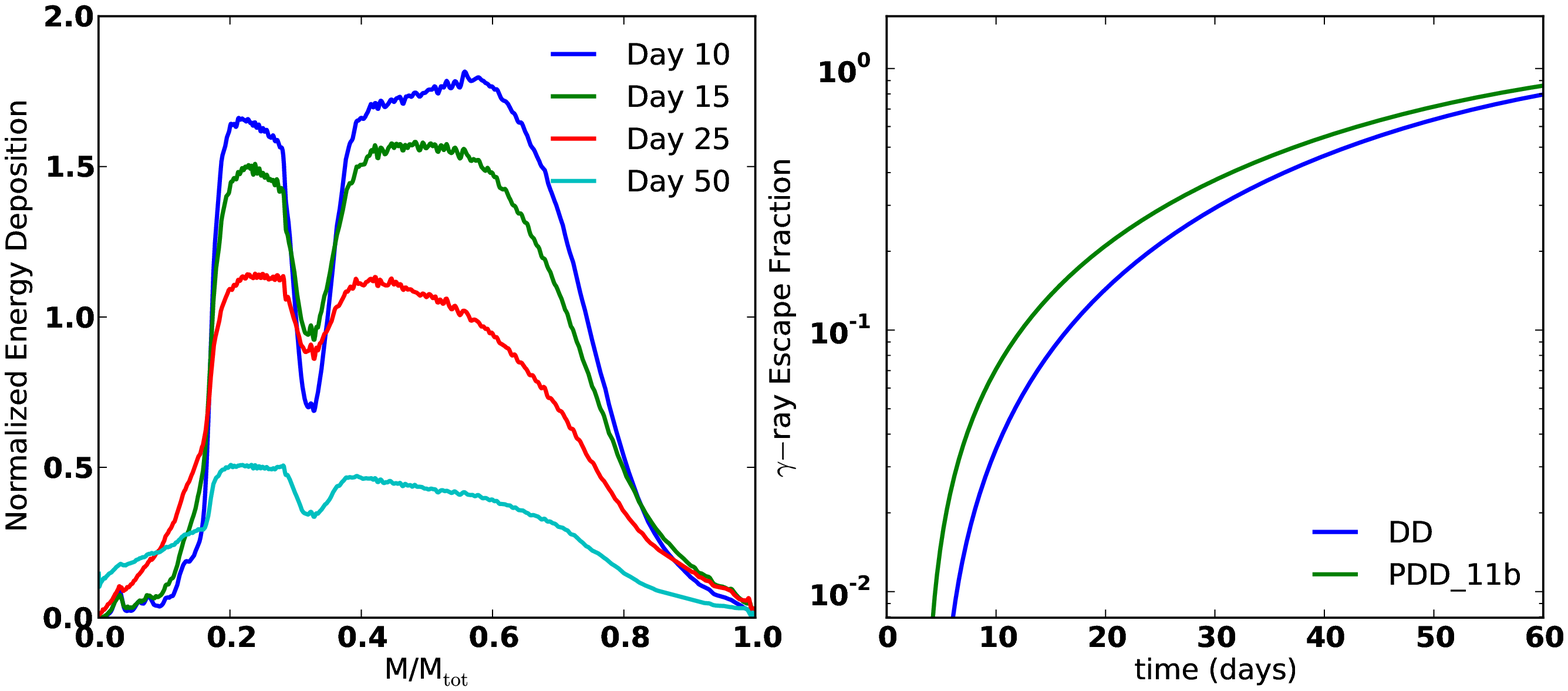}
\caption{We show the normalized energy deposition by \gamray and
  positrons in PDD\_11b (left), and the escape fraction for PDD\_11b
  and the reference model (right). }
\label{gam}
\end{center}
\end{figure}

Let us examine the formation of the light curves
and the comparison with \snooneay.  First, consider the energy
input by \gamray{s}, positrons, and adiabatic cooling due to
expansion. We contrast PDD\_11b with our reference model. Both have a
similar structure with respect to the chemical layering and, thus,
similar opacities.  Due to the higher explosion energy, the expansion
rate of the inner layers is larger by about 25--30\%. PDD\_11b
shows an increased energy loss due to expansion work.  The opacities
for $\gamma $'s depend only on the mass column height, $\tau_M $, and
the electron/nucleon ratio which is close to 0.5.  $\tau_M $ scales
as $v \times t^{-2}$ resulting in a 50\% lower optical depth due to
the higher velocity.  The normalized energy deposition and escape
fractions of \gamray{s} are shown in Fig.~\ref{gam}.  The high
expansion rate results in strong heating by \gamray{s} in the outer
region and a higher escape probability for \gamray{s}. Between 10 and
25 days, PDD\_11b has a an escape fraction larger than that of the
reference model  by about a factor of
2.

\begin{figure}[t]
\begin{center}
\includegraphics[width=\textwidth]{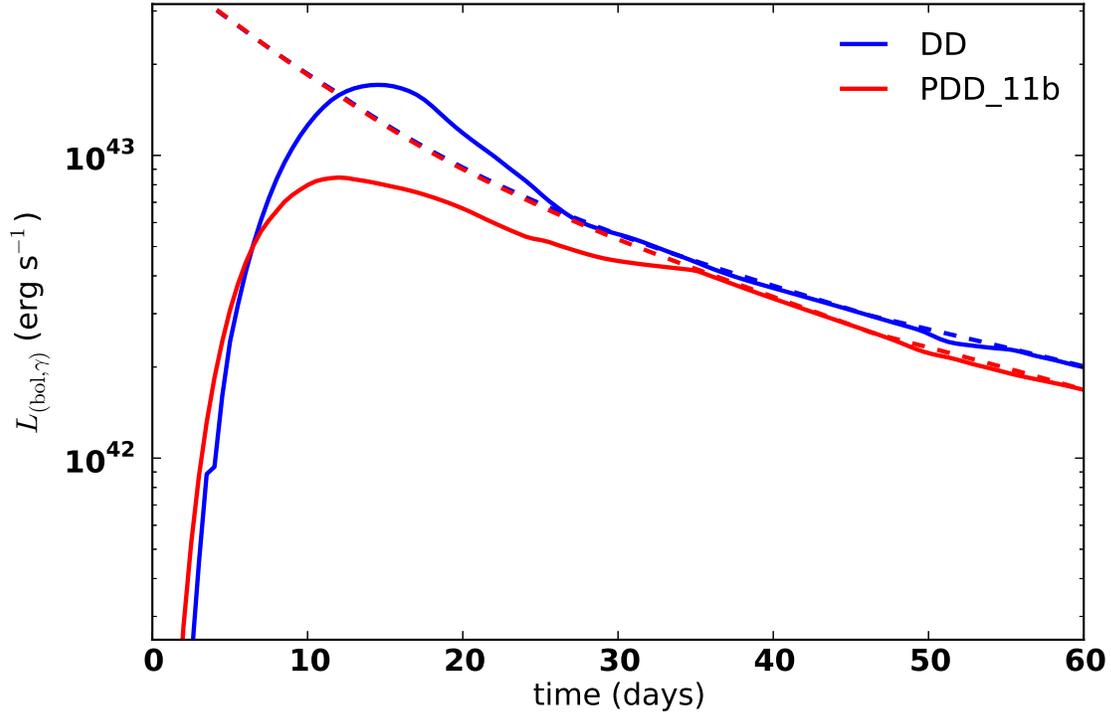}
\caption{ We give the instantaneous deposition rate due to
  radioactive decay (dashed) and bolometric luminosity (solid) for PDD\_11b, a pulsating model with a carbon rich
  core, and a ``classical'' DD model. The light curve of the DD model
  has been uniformly shifted by -0.037 dex and the light curve and \gamray
  deposition of the PDD\_11b model have been uniformily shifted by +0.05
  dex to better illustrate
  the variation of the gradients.}
\label{fig:sn2001ay_lc_bol}
\end{center}
\end{figure}

In Fig.~\ref{fig:sn2001ay_lc_bol}, we give a comparison between light curves of
PDD\_11b and the reference model.  PDD\_11b rises faster
and is brighter at early times. The higher escape of \gamray{s} and
the increased expansion work leads to $\alpha < 1$ and therefore, a
lower peak relative to the rate of instantaneous \gamray input.  The second
ingredient reducing the value of  $\alpha$ is the opacity difference
between optical photons and \gamray{s}. They are $0.1-0.2$ and $1/35
~\mathrm{cm{^2}~g^{-1}}$, respectively. \gamray{s} leak into the center
where the energy is trapped without contributing to the light curve at
maximum (Fig.~\ref{gam}).

We therefore have a deficit in luminosity (with respect to the
instantaneous \gamray deposition rate), and the optical light curves
approaches $\dot{E}_\gamma$ ``from below''.  The same opacity
mechanism responsible for the regular brightness-decline relation
results in an ``anti-Phillips''-relation! More \nni and with it,
larger opacities will lead to steeper declines.

\subsubsection{Comparison to \snooneay}

\begin{figure}[t]
\begin{center}
\includegraphics[height=.7\textheight,angle=-90]{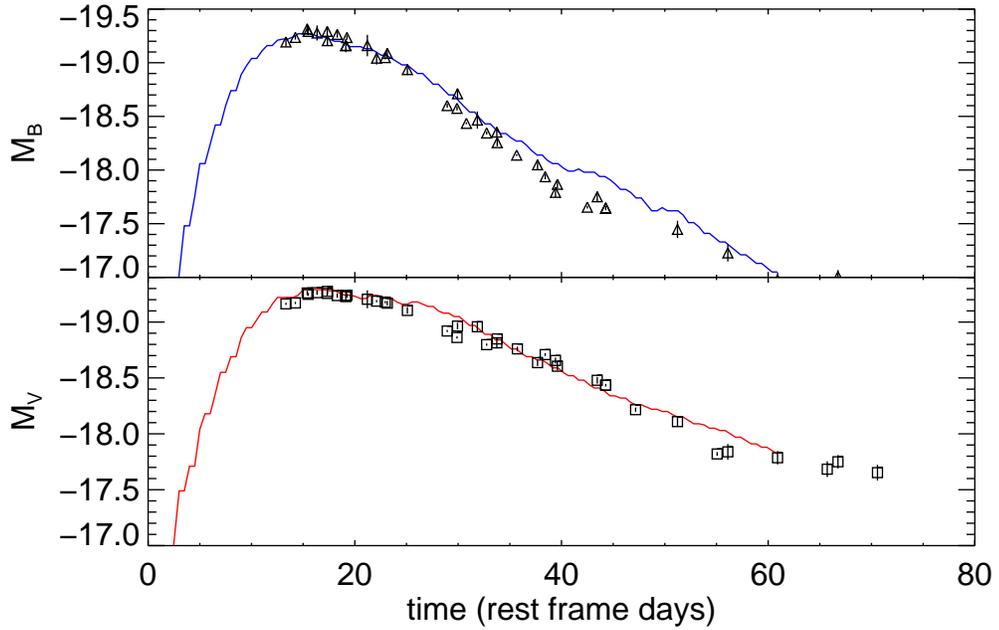}
\caption{ $B$ and $V$ LCs of \snooneay \citep{krisc01ay11} in
  comparison with theory. The comparison between \snooneay
  and PDD\_11b in $B$ (upper) and $V$ (lower) as a function
  of time since maximum light in the $V$-band. The model light curves
  have been corrected for reddening and redshift.  We assume a distance
  modulus m-M of $35.55$~mag, and following \citet{krisc01ay11}, we
  take $A_V = 0.253$ and $A_B = 0.35$.
  The data includes the photometric errors.}
\label{fig:sn2001ay_lc}
\end{center}
\end{figure}

In Fig.~\ref{fig:sn2001ay_lc}, the theoretical LCs in $B$ and $V$ are
compared with the observations.  The agreement is reasonable, and they
meet the brightness limit imposed by the early non-detection in $R$.
The theoretical $B$ and $V$ have been corrected for the redshift $z$ of
the host galaxy.  As discussed in the introduction, the distance
modulus of the host galaxy is $35.5 \pm 0.1$~mag, and the galactic
foreground extinction $E(B-V)=0.026$~mag.  \citet{krisc01ay11} use a
reddening of the host galaxy of $E(B-V)=0.072$~mag with an $R_V$ of
3.1 and 2.4 for the Galaxy and host galaxy, respectively, giving
$A_V=0.253$~mag.  Using our theoretical color, $(\bmv )_{\mathrm{max}}=0.0 $~mag,
and from an an optimized fit, we find the host galaxy reddening
$E(B-V)=0.02$~mag with a global $R_V=3.1$ for a total extinction
$A_V=0.144$~mag.

\subsection{Spectral Analysis}
 
Based on the explosion models, \gamray transport, and light curves,
we analyzed the spectrum of \snooneay at maximum light using the
NLTE-code \phx (see Appendix).

We use the density, abundance structure, and \gamray and
positron deposition given by the explosion models and light curve
calculations.  For the reddening and distance modulus, we use the
values found from our LC fit. For the redshift, we use
$z=0.030244$. Converged models required 256 optical depth points.

\begin{figure}
\includegraphics[width=0.60\textwidth,angle=270]{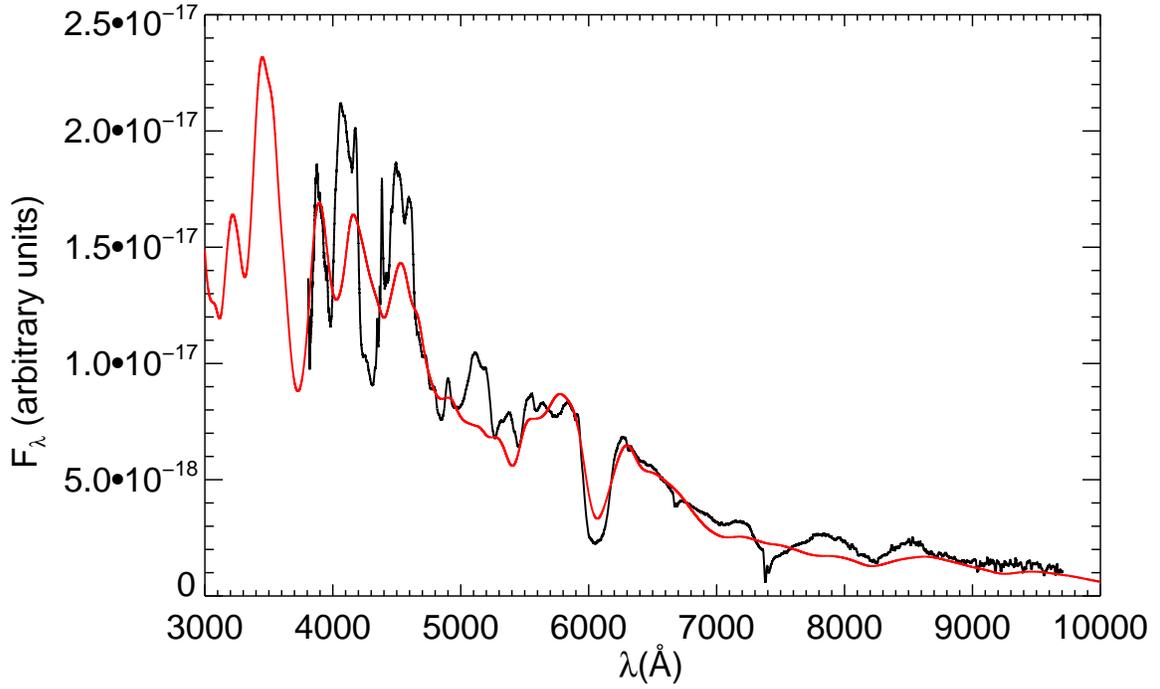}
\caption{Synthetic spectrum at day 16 of PDD\_11b in comparison with
  maximum spectrum of \snooneay. The density and abundance structure
  and the energy input by radioactive decay are taken from the
  explosion and light curve calculation
  (Figs.~\ref{fig:abundance}, \ref{gam}, and \ref{fig:sn2001ay_lc}). Roughly consistent
  with the light curve, $M_V $ was taken to be -19.07 mag, we have
  assumed standard reddening $R_V=3.1$ and E(B-V) = 0.096.}
\label{fig:syn_spectrum_cons}
\end{figure}

For a consistent absolute $M_V$ magnitude of $-19.1 $~mag, the resulting
synthetic spectrum is compared to the spectrum observed at maximum
light in Figure~\ref{fig:syn_spectrum_cons}.  The continuum colors are
well reproduced. The synthetic value of \bmv = -0.06~mag,
roughly  consistent with $0 $~mag of \snooneay. In order to avoid the
complications of two dust populations that was assumed by
\citet{krisc01ay11}, we have taken E(B-V) = 0.096 and $R_V = 3.1$, so
our reddening is a bit ``bluer'' that was assumed in the light curve
comparison, but it is well within the 0.1 magnitude error of the photometry.

The spectrum is dominated by single ionized lines of S~II, Si~II, Ca~II,
Fe~II, Co~II as well as blends of doubly ionized species in the
blue. The synthetic and observed spectra show good agreement.
The Doppler shifts of lines from elements undergoing incomplete oxygen
burning include  Si~II $\lambda 6355 $, Si~II $\lambda 5972 $ and
S~II $\lambda 5468 $ and $\lambda 5654$, and the Ca II H+K and the IR
triplet are reasonably well reproduced within
10, 20, and 30 \AA, respectively, which  corresponds to a
velocity shift of $500--1000 $~\kmps at the  measured
velocity of $14,400 $~\kmps. The strength of the absorption components
agree well. 
A large number of weak lines form the
quasi-continuum in the near IR.  The feature due to blends of Fe~II/Co~II/Ni~II which
could possibly be misidentified as due to C~II 6580 is reasonably strong, but
the feature does not agree with the absorption notch identified as C~II by
\citet{krisc01ay11}. However, the evidence for C~II claimed in
\citet{krisc01ay11} is stronger than warranted based on the analysis
that were done for that paper (R.~Thomas, personal communication).
 Some disagreement is evident.  The
S~II W at about 5000 \AA, blended with other lines, and the Si~II
$\lambda 5970$ line are too weak.  In the red, the  Ca IR
triplet is too weak, as is the O~I $\lambda\lambda 7773$ line, which
is however, severely contaminated by Telluric absorption. The lack of
O~I is an indication that the outer part of our model is not correct,
either in density structure or due to temperature effects. 

To probe importance of consistency and the sensitivity of the spectra,
we calculated a series of maximum light spectra which are under- and
over-luminous,  with $M_V$ = (-18.92,-19.23,-19.35,-19.45)~mag.

\begin{figure}
\includegraphics[width=0.60\textwidth,angle=270]{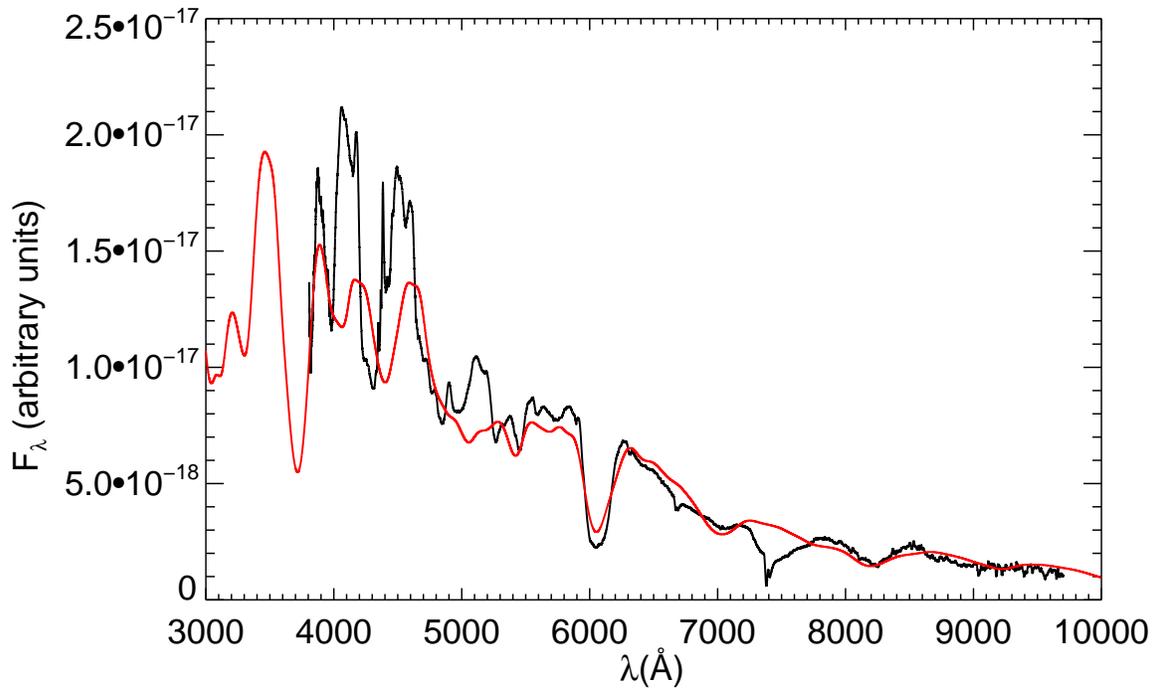}
\caption{Same as Fig.~\ref{fig:syn_spectrum_cons} but dimmer
   by 0.1 mag, i.e. $M_V$ of -18.92 mag. 
}
\label{fig:syn_spectrum_lt1}
\end{figure}

At +0.1 mag (Fig.~\ref{fig:syn_spectrum_lt1}, the \bmv color becomes
$0.03$~mag. The Si~II $\lambda 6355$ line is well-fit and the Ca~II IR
triplet is relatively well reproduced. In the blue, Co~II lines
dominate and a Co~II line at 7400~\AA\, appears. The feature at 6500 \AA\,
is weaker.
The Doppler shift of the
Si~II and S~II remain in good agreement because they are formed within the
region of incomplete oxygen burning which produced nearly constant
abundances due to  burning in quasi-statistical equilibrium (QSE) conditions
(Fig.~\ref{fig:abundance}).

\begin{figure}
\includegraphics[width=0.60\textwidth,angle=270]{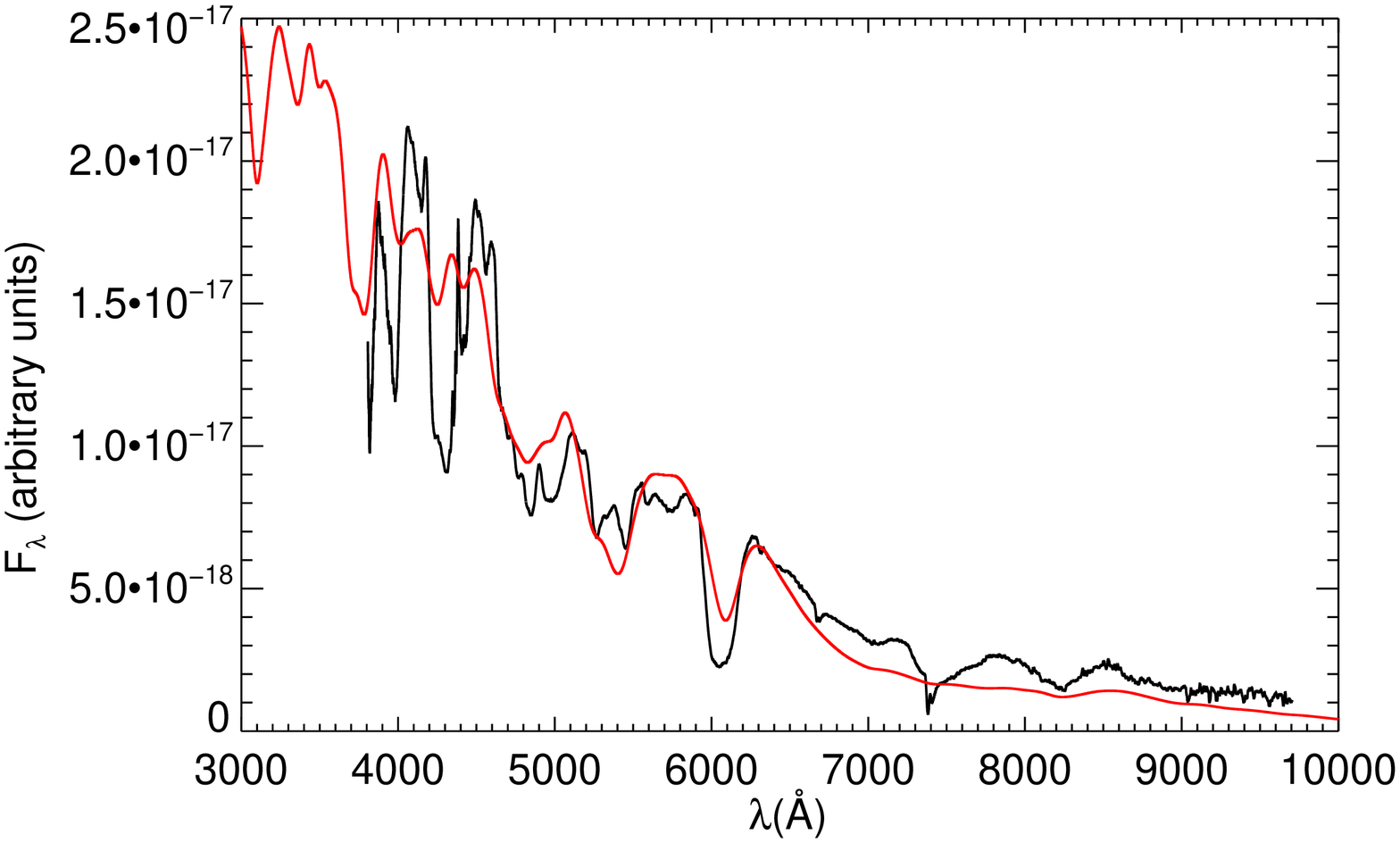}
\caption{Same as Fig.~\ref{fig:syn_spectrum_cons} but brighter
   by 0.15 mag, i.e. $M_V$ of -19.23 mag. 
}
\label{fig:syn_spectrum_lo1}
\end{figure}

At -0.15 mag (Fig.~\ref{fig:syn_spectrum_lo1}), 
\bmv equals $-0.11$~mag, but is clearly a
little bit too blue as can be seen  in the red part of the spectrum
However, the Si~II $\lambda 5970$
feature is too weak and S~II lines are still significantly too
weak. The Ca~II IR triplet is much too weak. In the blue, Fe~II lines
are stronger and the fit is somewhat better. This could imply that the
model should have had an initial metallicity somewhat higher, which
would create more Fe~II at the expense of Co~II (from radioactive nickel).

\begin{figure}
\includegraphics[width=0.60\textwidth,angle=270]{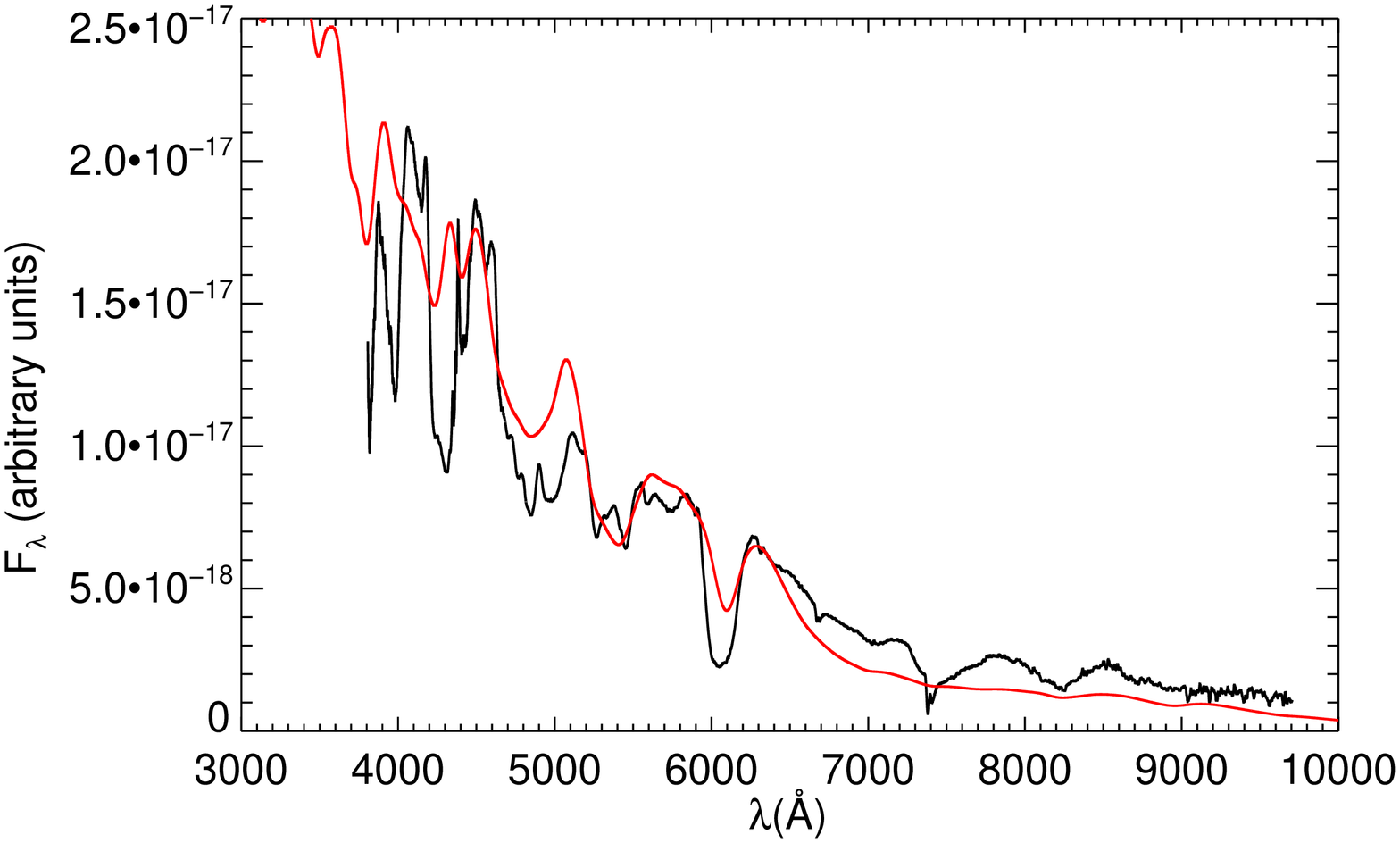}
\caption{Same as Fig.~\ref{fig:syn_spectrum_cons} but 
  brighter by 0.25 mag, i.e. $M_V$ of -19.35 mag.} 
\label{fig:syn_spectrum_lo2}
\end{figure}

At -0.25 mag (Fig.~\ref{fig:syn_spectrum_lo2}), the spectrum becomes
too blue, \bmv = -0.08 mag, and the UV flux becomes too large.  The
Si~II and S~II lines are still significantly too weak and the IR flux
is too low. The Ca H+K line is now too weak.

\begin{figure}
\includegraphics[width=0.60\textwidth,angle=270]{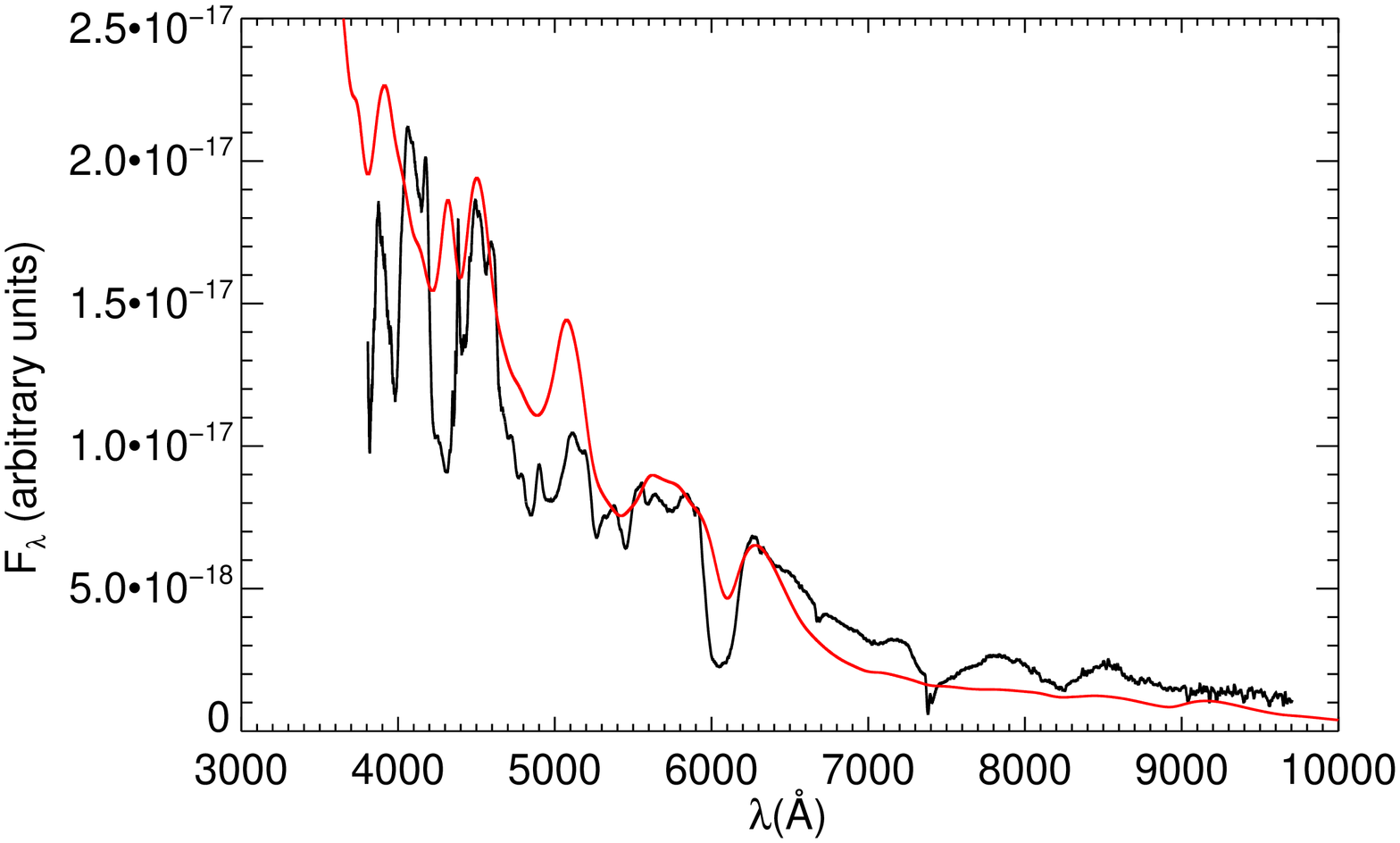}
\caption{Same as Fig.~\ref{fig:syn_spectrum_cons} but 
  brighter by 0.35 mag, i.e. $M_V$ of -19.45 mag.} 
\label{fig:syn_spectrum_lo3}
\end{figure}

At -0.35 mag (Fig.~\ref{fig:syn_spectrum_lo3}), the spectrum is much
too blue, \bmv = -0.07~mag and the 
Ca~II features are completely absent. The Si~II and S~II features are
now much too weak. 
   
\begin{figure}
\includegraphics[width=0.95\textwidth,angle=0]{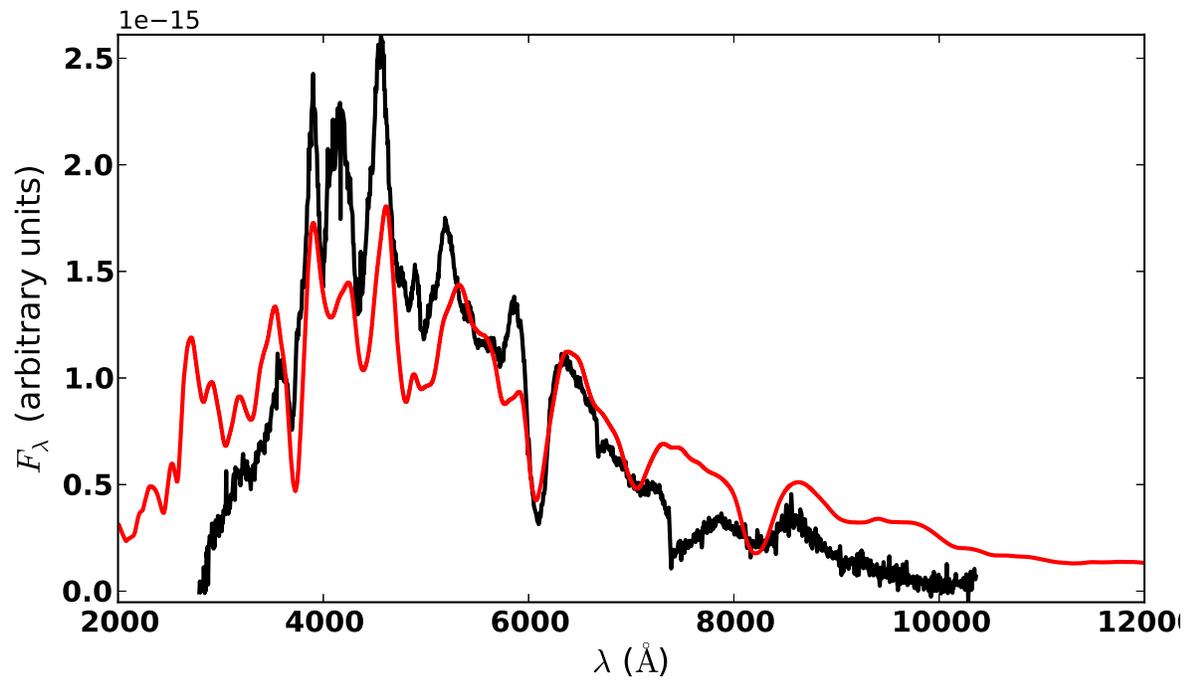}
\caption{Synthetic spectrum (red) for 23 days after explosion compared with
  the observed spectrum obtained on Apr 29, 2001, from both \emph{HST}
  and Keck. 
  $M_V$ is  -18.97 mag. and $B\,-V = 0.43$.} 
\label{fig:syn_spectrum_day23_l3}
\end{figure}

\begin{figure}
\includegraphics[width=0.95\textwidth,angle=0]{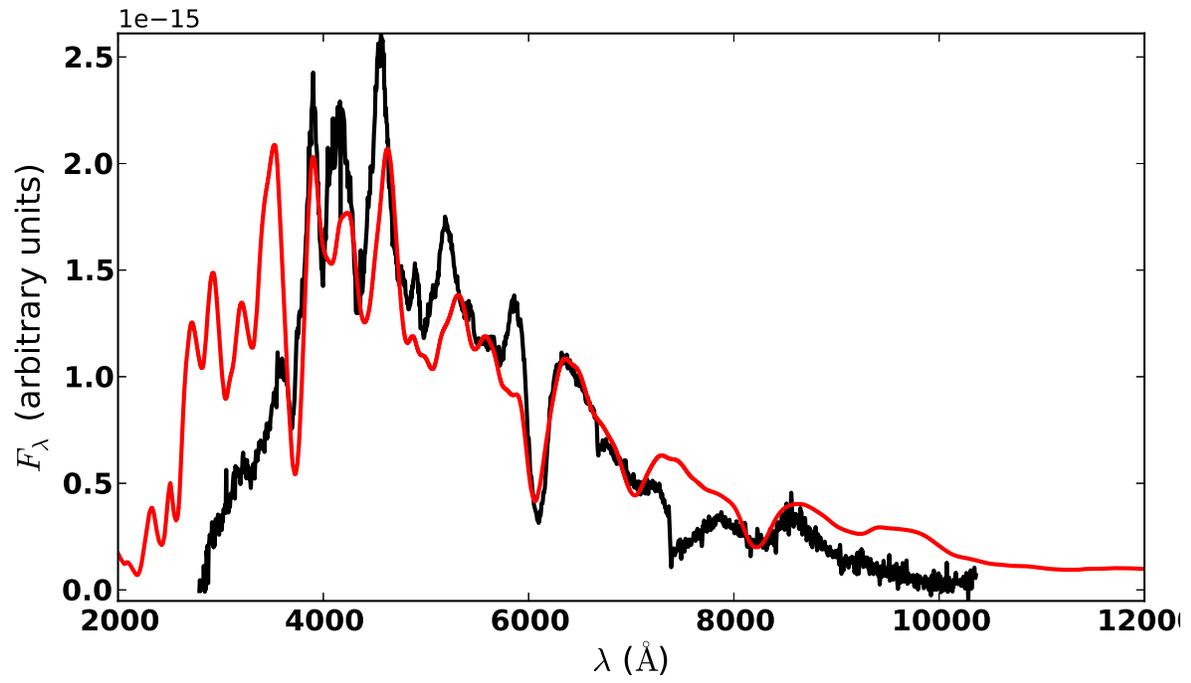}
\caption{Same as Figure~\ref{fig:syn_spectrum_day23_l3}, but 
  $M_V= -19.22$ mag. and $B\,-V = 0.22$.} 
\label{fig:syn_spectrum_day23_l1}
\end{figure}

\begin{figure}
\includegraphics[width=0.95\textwidth,angle=0]{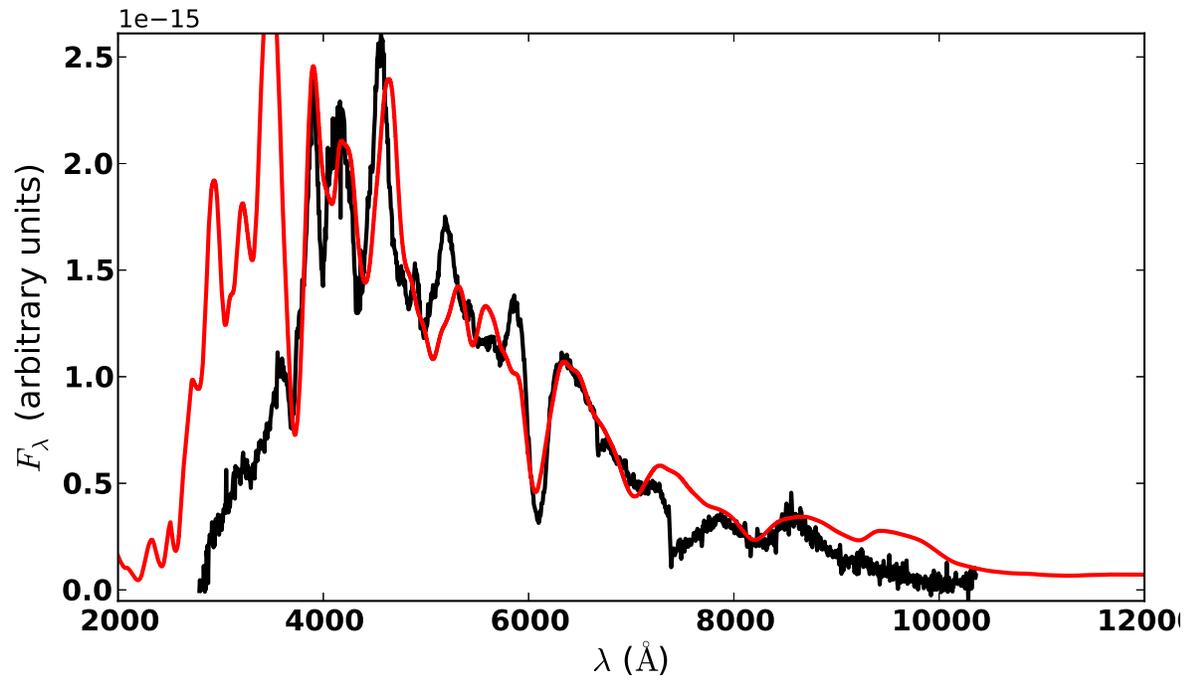}
\caption{Same as Figure~\ref{fig:syn_spectrum_day23_l3}, but 
  $M_V = -19.4$ mag. and $B\,-V = 0.10$.} 
\label{fig:syn_spectrum_day23_l5}
\end{figure}

Figure~\ref{fig:syn_spectrum_day23_l3} shows the model fit to the
observed \emph{HST}+Keck spectrum of April 29. The Si~II line is very nicely fit,
the S~II ``W'' is well fit, although the blue line is a bit stronger
in the model than in the observed spectrum. The rest of the blue is
reasonably well fit, although the flux is about a factor of 2 too high in the
UV, which may be an indication of the need for a  higher metallicity,
which would produce more line blanketing and reduce the UV flux. Both the Ca
H+K and IR triplet are reasonably well fit, and the Co~II + Fe~II + Mn~II
feature at about 7350 \AA\ is a bit too strong. The O~I 
$\lambda\lambda 7773$ line is much too
weak. Figure~\ref{fig:syn_spectrum_day23_l1} shows the model fit for a
somewhat brighter model $M_V = -19.22$ and $B\,-V = 0.22$, still a bit
too red, but the line features fit similarly well to the redder case.
Figure~\ref{fig:syn_spectrum_day23_l5} shows the model fit for a
somewhat brighter model $M_V = -19.40$ and $B\,-V = 0.10$, a bit too
bright, but about the right color. Again, the too high UV and the weak
O~I line are indications that the outer structure of our model is not
quite correct, which could be due to primordial metallicity or our
assumptions about mixing in the PDD.

\begin{figure}
\includegraphics[width=0.95\textwidth,angle=0]{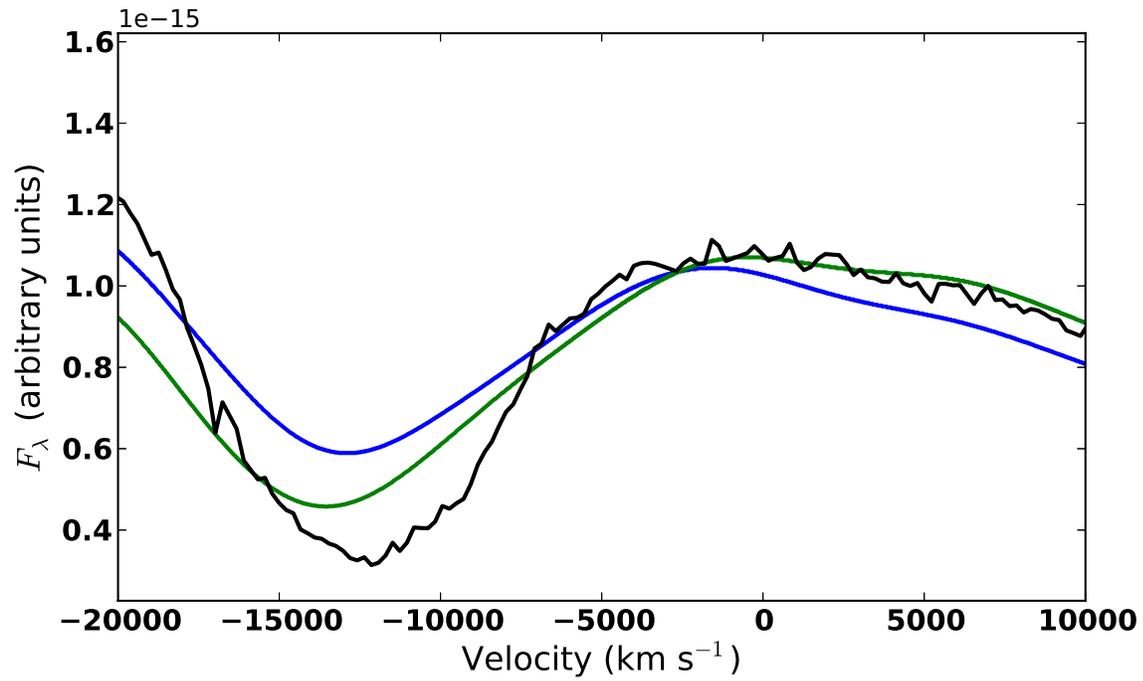}
\caption{The Si~II $\lambda 6355$ feature for 23 days after explosion compared with
  the observed spectrum obtained on Apr 29, 2001 (black) with the model
  shown in Fig.~\ref{fig:syn_spectrum_day23_l5} (green) and a somewhat hotter
  model (blue). The blueward shift in the absorption is clearly a
  temperature effect and is also due to blending. }
\label{fig:si_day23_l5_l4}
\end{figure}

Obviously, spectral synthesis is a very sensitive tool to probe the
structure and luminosity on an level of $\Delta M_V \approx
0.1$~mag. Variations in excess of $\Delta M_V \approx 0.2$ lead to spectral
fits which would lead to a rejection of even a valid, underlying
explosion model. We note that our best fits give a ``correct'' velocity
shift of e.g. the Si~II feature because, in PDD\_11b, it is formed in
the region of QSE for the Si/S group. Consistent treatment of
explosion, light
curves, and spectral formation is  
important. However, spectra lines are sensitive to both temperature
effects and line blending and even a strong line like Si~II $\lambda
6355$ is not unblended. One can see a small blueward shift between
Figs.~\ref{fig:syn_spectrum_cons} and
\ref{fig:syn_spectrum_day23_l3}. Figure~\ref{fig:si_day23_l5_l4} shows
the Si~II $\lambda 6355$ feature in velocity space, where two models
at the same epoch (Apr 29) are plotted along with the data. The cooler
model has an absorption minimum that is bluer than that of the hotter
model and both are a bit bluer than the observed spectrum, although
the noise in the spectrum makes a determination of the minimum a bit difficult.

In turn, light curves are mostly sensitive to the inner region and the
large number of weak lines.  Both in the LC and spectral calculation,
\bmv is consistent, namely 0 and -0.06 mag. Only spectra show
that the Si~II line is too weak.

As discussed above, spherical models inherently suppress mixing by
instabilities due to the deflagration phase at the inner
layers, and the interaction between shells and the outer layers 
during the acceleration phase. 

Spectral analysis requires underlying models which are consistent
including the luminosity. Then, however, they allow the study of secondary
effects of individual spectral features.

\section{Discussion and Conclusion}

The peculiar Type Ia  \snooneay is an important milestone for our
understanding of the explosion, light curves, and spectra.

Even with its extremely unusual light curve shape and spectral features,
we showed that it still can be accommodated within the framework of
the explosion of a Chandrasekhar mass white dwarf.

Although it does not follow the \dmphil--relation, the light curve can be
reproduced within the physics of normal SNe~Ia, and in the framework
of pulsating delayed detonation models. In our models, the key
difference is a high, 80\% carbon mass fraction, rather than the
15--20\% carbon mass fraction 
usual for stellar central He-burning. The size of the carbon rich core
is 0.56~\msol. The excess of carbon coupled
with a lower central density of the initial WD results in an increase
of the final kinetic energy by about 40\% and in turn, shifts the
distribution of \nni to larger velocities.  These modifications are
responsible for both the LC characteristics and the broad spectral
features.

The peculiar light curve shape is a consequence of the shift of the \nni
distribution in velocity space and higher rate of central
expansion. Transport effects 
of \gamray{s} become more important and the escape probability is
increased. This leads to a fast, early rise and a value of $\alpha <1 $, where
$\alpha $ is the ratio between the luminosity and the instantaneous energy
generation rate by radioactive decays at maximum.  In normal SNe~Ia,
the \dmphil--relation can be understood as a consequence of the temperature
dependence of the opacity in combination with $\alpha > 1$. As
discussed in \S~\ref{lc_dis}, more \nni causes higher
luminosities and temperatures. Higher temperatures lead to larger
opacities and therefore, a slower drop of the luminosity from the
typical value of  $\approx 1.2 \times $ the instantaneous energy input rate
at maximum, to the
instantaneous energy input rate at later times.  Accordingly, for \snooneay with
$\alpha \approx 0.55$, the light curve approaches the instantaneous
energy input rate
from below which explains the unusually slow decline.  Within our
models and as a corollary, we expect that there exists  a sub-class of
supernovae which 
obeys an anti-Phillips relation.  As discussed below, this sub-class should be
rare.

The unusually broad spectral features can be understood by the overall
shift of the overall element pattern to higher
velocities. However, this pattern remains similar to that found in  normal SNe~Ia
(Fig.~\ref{fig:abundance}). We demonstrated the power and sensitivity of
spectral analysis. High sensitivity demands a consistent treatment with
the explosion and light curve models. Otherwise, even valid models may
be rejected because of poor fits. However, since a good overall fit
is achieved with a consistent treatment, individual features are a
powerful tool to study details of the explosion physics.  

We have shown that limiting the mixing during the pulsation to small
scales produces a very similar abundance pattern to the one produced in standard
DD-models which have been shown to reproduce the observables for the
majority of SNe~Ia. PDD may be a promising mechanism for the DDT.

While our model fits the basic observed features of \snooneay, we do
not mean to imply that other models are not possible. In particular a
DD model 
including 3-D effects as discussed above with a high central C/O ratio may also be a viable model
and there may exist other models that fit the observed
trends. Pursuing other explanations is beyond the scope of the present work.

Our model agrees reasonably well with the observations
(Fig.~\ref{fig:sn2001ay_lc}) but the high carbon abundance in the progenitor
poses a challenge. The central region of the progenitor originates
from central He burning in stars with less than 7--8 \msol
\citep{BI80}.  During the early stages of central He burning, high carbon
abundances are produced by $^4\mathrm{He}(2 \alpha,\gamma)^{12}C$ burning.
With time, the helium abundance is reduced in the core. Then,
$^{12}C(\alpha, \gamma)^{16}O $ depletes $^{12}C$ to a typical value
of 10--25\% \citep{umeda99,DHS01a}. The final amount of
$^{12}C$/$^{16}O$ depends on the mass of the progenitor, and the
$^{12}C(\alpha, \gamma)^{16}O $ rate
 \citep{Buch97}, and the amount chemical mixing assumed
\citep{Castellani85,Caputo89,Renzini88,DHS01a}.  Increased chemical
overshooting (or semiconvection) prolongs the phase of burning under helium depleted
conditions, leading to a lower value of C/O. Although chemical
overshooting prescriptions
vary widely between various  groups studying stellar
evolution, the final outcome is a C-poor mixture.  To reach high
central carbon
abundances, burning under He-depleted conditions must be
avoided \citep[see][and references therein]{SDIP03}. \citet{SDIP03}
found that they could increase the central carbon abundance somewhat
by increasing mechanical overshooting.
As a possible solution, we propose a common envelope phase
with very strong mixing induced by a compact secondary such as a brown
dwarf or planet. Common envelope evolution is generally assumed to be
responsible for the formation of close binaries leading to cataclysmic
variables, X-ray binaries, and supersoft-X-ray sources
\citep{LS88,RT08}.  Moreover, close binary systems with
planet or brown dwarfs have recently been detected
\citep{Neu07,Hessman11}. Unfortunately, there are no systematic
studies which allow one to estimate the amount of mixing. Detailed,
numerical studies are few, and analytic models are insufficient to
quantify the amount of chemical mixing \citep{MM79}.  One
observational clue may 
be SN~1987A, which is believed to be the result of a common
envelope evolution \citep{Pod90}. Blue progenitors may also be a
result of low metallicity \citep{BrTr82,Chieffi03}, and very few
SN1987A-like events have been found
\citep{pasto09E_12,Taddia12}. \citet{pasto09E_12} estimate that SN~1987A-likes
represent $\sim 1.5-4$\% of SNe~II; however, this is likely an extreme
upper limit to our scenario, since the preponderance of these objects
may be just from higher mass compact stars, rather than from systems
in binaries that have undergone common envelope evolution. Similarly,
one can use observations of white dwarfs to provide another
estimate. The number of white dwarfs with sub-stellar companions is
$\la 0.5$\% \citep{FBZ05}, and the number of white dwarfs with debris
disks from tidally disrupting minor planets is $1-3$\% \citep{FJZ09}
Therefore, our scenario should be quite rare, representing $0.05 -
0.5$\% of all SNe Ia. 

Finally, we want to mention the long list of limitations of our
studies: We have suggested the existence of a rare sub-class of SNe~Ia
which should obey an 
anti-Phillips relation. It is up to future, systematic surveys, such as
LSST to find a sufficiently large statistical sample. Note, that this
sub-class may be hard to separate at the bright end of SNe~Ia because
the decline rates for both normal and ``\snooneay-like'' supernovae are
low and similar.  Chemical mixing during the common envelope phase
needs to be studied in detail.  Our model for \snooneay requires a
shift of the \nni distribution and a central hole in \nni.  Although
verified for a number of normal SNe~Ia, late time spectra for a
\snooneay-like are required to confirm our
assumptions of little mixing.  Detailed time-series of early time
spectra may help to probe whether PDDs provide a common mechanism
for the transition from deflagration to detonation. 
Detailed 3D models of the pulsation phase are required to test
and quantify possible mixing.

\acknowledgments The work presented in this paper has been carried out
within the NSF project ``Collaborative research: Three-Dimensional
Simulations of Type Ia Supernovae: Constraining Models with
Observations'' whose goal is is to test and constrain the physics of
supernovae by observations and improve SNe~Ia as tools for high
precision cosmology. The project involves The University of Chicago
(AST-0709181), the University of Oklahoma (AST-0707704), Florida State
University (AST-0708855), Texas A\&M (AST-0708873), The University of
Chile in Santiago, and the Las Campanas Observatory, Chile.  This
research was also supported, in part, by the NSF grant AST-0703902 to
PAH.  The work of EB was also supported in part by SFB 676, GRK 1354
from the DFG, and US DOE Grant DE-FG02-07ER41517. 
ID has been supported in part by the Spanish Ministry
of Science and Innovation project AYA2008-04211-C02-02 (ID).
  Support for Program
number HST-GO-12298.05-A was provided by NASA through a grant from the
Space Telescope Science Institute, which is operated by the
Association of Universities for Research in Astronomy, Incorporated,
under NASA contract NAS5-26555.  This research used resources of the
National Energy Research Scientific Computing Center (NERSC), which is
supported by the Office of Science of the U.S.  Department of Energy
under Contract No.  DE-AC02-05CH11231; and the H\"ochstleistungs
Rechenzentrum Nord (HLRN).  We thank both these institutions for a
generous allocation of computer time.

\clearpage

\bibliography{apj-jour,mystrings,refs,baron,sn1bc,sn1a,sn87a,snii,stars,rte,cosmology,gals,agn,atomdata,crossrefs}

\begin{appendix}
\label{apx:a}

\section{Brief Description of numerical Methods}

\subsection{Explosion}
We have calculated explosion models and light curves using the
one-dimensional radiation-hydro code HYDRA using computational modules
for spherical geometry \citep{h95,h03a,h03b,h09}.
 We solve the hydrodynamical equations explicitly by
the piecewise parabolic method on 910 depth points \citep{CW_PPM84}.
Because a simple $\alpha $ network is insufficient to describe in
sufficient detail the chemical boundary. Nuclear burning is taken into
account using an extended network of 218 isotopes from n, p to
$^{74}\mathrm{Kr}$ \citep[and references therein]{thielemann96,hix96,hwt98}. The
propagation of the nuclear burning front is given by the velocity of
sound behind the burning front in the case of a detonation wave. We
use the parametrization as described in \citet{DH00}.  For a
deflagration front at distance $r_{\mathrm{burn}}$ from the center, we assume
that the burning velocity is given by $v_{\mathrm{burn}}=\mathrm{max}(v_{t},
v_{l})$, where $v_{l}$ and $v_{t}$ are the laminar and turbulent
velocities with
$$ v_{t}= ~\sqrt{\alpha_{T} ~ g~L_f},~
\eqno{[1]} $$
 with~
$$\alpha_T ={(\alpha-1)/( \alpha +1})$$ ~and ~
$$\alpha ={\rho^+(r_{\rm burn})/ \rho^-(r_{\rm burn})}.$$
\noindent
Here $\alpha _T$ is the Atwood number, $L_f$ is the characteristic
length scale, and $\rho^+$ and $\rho^-$ are the densities in front of
and behind the burning front, respectively. The quantities $\alpha$ and $L_f$ 
are directly taken from the hydrodynamical model at the location of the burning 
front and we take $L_f=r_{\mathrm{burn}}(t)$. The transition density
is treated as a  
free parameter. Table~\ref{tab:yield} gives the yields of the stable
elements as well as the total amount of \nni produced in the explosion.

\begin{deluxetable}{rr}
\tablecolumns{2}
\tablewidth{0pc}
\tablecaption{Element Yields \label{tab:yield}}
\tablehead{
\colhead{Element}    &  \colhead{Yield (\msol)}
}
\startdata
He & 0.01\\
C  & 0.021\\
O  & 0.043\\
Ne & $3\times 10^{-3}$\\
Na & $7.8\times 10^{-5}$\\
Mg & $1\times 10^{-3}$\\
Si & 0.182\\
P  & $9\times 10^{-5}$\\
S  & 0.104\\
Cl & $4.2\times 10^{-5}$\\
Ar & 0.022\\
K  & $4.8\times 10^{-5}$\\
Ca & 0.022\\
Sc & $3.1\times 10^{-7}$\\
Ti & $1.5\times 10^{-3}$\\
V  & 0.022\\
Cr & 0.106\\
Mn & 0.024\\
Fe & 0.661\\
Co & $1.1\times 10^{-3}$\\
Ni & 0.146\\
\nni & 0.515\\
\enddata
\tablecomments{The yield of the stable elements at time
  infinity along with the total mass of \nni produced in the explosion.}
\end{deluxetable}

\subsection{Light Curves}
From these explosion models, the subsequent expansion, bolometric and
broad band light curves (LC) are calculated using monochromatic
radiation transport via the Eddington Tensor method. Both the Eddington
tensor and the \gamray transport are calculated via Monte Carlo
as in the 3D case.  We include scattering, free-free, and bound-free
opacities, and include the line transitions in the Sobolev
approximation. For several elements, including C, O, Mg, Si, S, Ca,
Fe, Co, and Ni
we solve the statistical equations for the three main ionization
stages using detailed atomic models with 10--50 super-levels with
$\approx 12,000 $ transitions using the databases of \citet{kurucz02}
and \citet{topbase}.  The 
levels closest to the continuum and of other elements are treated in
local thermodynamical equilibrium with about $10^6 $ line transitions.

\subsection{Spectra}
The calculations were performed using the multi-purpose stellar
atmospheres program \phxO~{version \tt 16}
\citep{hbjcam99,bhpar298,hbapara97,phhnovetal97,phhnovfe296}. Version
16 incorporates many changes over previous versions used for supernova
modeling \citep{bbh07,bbbh06} including many more species in the
equation of state (83 versus 40), twice as many atomic lines, and many
more species treated in full NLTE and an improved equation of state.
\phxO solves the radiative transfer equation along characteristic rays
in spherical symmetry including all special relativistic effects.  The
non-LTE (NLTE) rate equations for many ionization states are solved
including the effects of ionization due to non-thermal electrons from
the \gamray{s} produced by the  radiative decay of $^{56}$Ni, which
is produced in the
supernova explosion.  The atoms and ions calculated in NLTE are: 
  He~I--II, C~I--III, O~I--III, Ne~I, Na~I--II, Mg~I--III,
  Si~I--III, S~I--III, Ca~II, Ti~II, Cr~I--III, Mn~I--III, Fe~I--III,
  Co~I--III, and Ni~I--III. 
 These are all the
elements whose features make important contributions to the observed
spectral features in SNe~Ia. \gamray deposition was that calculated by
the light curve. 

Each model atom includes primary NLTE transitions, which are used to
calculate the level populations and opacity, and weaker secondary LTE
transitions which are included in the opacity and implicitly
affect the rate equations via their effect on the solution to the
transport equation \citep{hbjcam99}.  In addition to the NLTE
transitions, all other LTE line opacities for atomic species not
treated in NLTE are treated with the equivalent two-level atom source
function, using a thermalization parameter, $\alpha =0.10$
\citep{snefe296}.  The 
atmospheres are iterated to energy balance in the co-moving frame;
while we neglect the explicit effects of time dependence in the
radiation transport equation, we do implicitly include these effects,
via explicitly including $p\,dV$ work and the rate of gamma-ray
deposition in the generalized 
equation of radiative equilibrium and in the rate equations for the
NLTE populations.

The outer boundary condition is the total bolometric luminosity in the
observer's frame. The inner boundary condition is that the flux at the
innermost zone ($v=700$~\kmps) is given by the diffusion
equation. Converged models required 256 optical depth points to
correctly obtain the Si~II $\lambda 6355$ profile.

\phx\ and HYDRA have been well tested and compared on SNe~Ia
\citep{nug1a95,nugseq95,nughydro97,l94d01,bbbh06,h05,hwt98,h02} and, in particular
compared with observed light curves and spectra of SN~1994D and \snooneay.

\end{appendix}

\clearpage 

\clearpage

\end{document}

\begin{deluxetable}{rr}
\tablecolumns{2}
\tablewidth{0pc}
\tablecaption{Element Yields}
\tablehead{
\colhead{Element}    &  \colhead{Yield (\msol)}
}
\startdata
He & 0.0100356119486\\
C  & 0.0209326629264\\
O  & 0.0429375678307\\
Ne & 0.00288511376796\\
Na & 7.88757262215e-05\\
Mg & 0.00103969161211\\
Si & 0.182368092549\\
P  & 8.94895584357e-05\\
S  & 0.104252173297\\
Cl & 4.16962411008e-05\\
Ar & 0.0226407029546\\
K  & 4.85489559686e-05\\
Ca & 0.0216462887596\\
Sc & 3.12299182676e-07\\
Ti & 0.00153953319586\\
V  & 0.0216118799742\\
Cr & 0.106570321626\\
Mn & 0.0244316418502\\
Fe & 0.661171962453\\
Co & 0.00107274616488\\
Ni & 0.146201817982\\
\nni & 0.515206745294
\enddata
\end{deluxetable}